\documentclass{article}

\usepackage{graphicx}

\newcommand{\bel}[1]{\begin{equation}\label{#1}}

\newcommand{\be}{\begin{equation}}

\newcommand{\ba}{\begin{eqnarray}}
\newcommand{\ea}{\end{eqnarray}}
\newcommand{\rf}[1]{(\ref{#1})}
\newcommand{\bi}{\bibitem}
\newcommand{\qe}{\end{equation}}

\begin{document}

\title{Evolution of network structure by temporal learning}

\author{J\"urgen Jost$^{1,*}$ and Kiran M. Kolwankar$^{1,2}$}

\maketitle

\begin{center}
\emph{$^1$ Max Planck Institute for Mathematics in the Sciences, Inselstrasse
22, D-04103 Leipzig, Germany \\
$^2$ Department of Physics, Ramniranjan Jhunjhunwala College, Ghatkopar (W), Mumbai 400 086, India}
\end{center}

\begin{abstract}
We study the effect of learning dynamics on  network topology.
A network of discrete dynamical systems is considered for
this purpose and the coupling strengths are made to evolve
according to a temporal learning rule that is based  on the paradigm of spike-time-dependent
plasticity. This incorporates necessary competition between
different edges. The final network we obtain is robust and has a broad degree distribution.
\end{abstract}

\vspace{0.5in}


\vspace{0.5in}
\noindent
$^*$Corresponding Author.

\noindent
Email: jjost@mis.mpg.de
Phone: +49-341-99 59 550   Fax: +49-341-99 59 555
\newpage

\section{Introduction}
In recent years, several paradigms have been advanced for 
 understanding the structural properties of
 networks in different domains, and certain rather universal features
 have been identified, like the small world property or a power law
 degree distribution, ~\cite{WS,BA}. In this context, also some
 principles for the generation of networks have been proposed that can be
 captured by slogans like ``the rich get richer''\cite{Sim,BA} or ``make friends
 with the friends of your friends''\cite{JJ2}. We here pursue a somewhat different approach, namely the modification of an
existing network by dynamical rules, but again with a view towards
universal features. Our model originates from the observation of the temporal development
of neuronal networks on the basis of the interplay between a fast
activity dynamics at the neurons, the nodes in the network, and a slow
learning dynamics affecting the strengths of the synapses, the edge
weights of the network, see \cite{DA} as a general reference.  Our aim here is to explore the possibility
that some rather universal features of a learning process operating on
and modifying a given edge  can lead to robust network
structures in the case of coupled chaotic oscillators. The neuronal networks in the brain consist of neurons connected to
each other via directed synapses that transmit electrical signals
between neurons. The potential of a neuron thus changes on a fast time scale 
depending on the inputs from other neurons. The strength of a synapse determines the efficiency
of this synapse to transmit a signal.
The dynamics of the strength of the synapses which is interpreted as
learning is much slower than the
dynamics of the neurons, see e.g. \cite{Sh}. 

It is well-known that just after  birth the brain has a very
dense population of synaptic connection and, as time goes on, 
most of the connections are pruned (see, for example,~\cite{Bis}). This obviously happens as
a result of learning. This  leads to the question
about the role of  learning dynamics in deciding the final network
structure.

\section{Model}

We wish to explore whether learning dynamics can lead to the pruning of a
densely connected network and result in a robust sparsely
connected network. We are not focussing on neurobiological details,
but rather on general principles. In order that a dynamical network
generate nontrivial behavior, we need some interplay of enhancing and
reversing features, at the level of the dynamics of the individual
elements or as a structural feature of the network connections. In
typical neural network paradigms, the latter is
achieved by the presence of both excitatory and inhibitory synapses,
while the dynamics of an individual formal neuron could be defined by
a monotonically increasing function of its inputs,  e.g. by a sigmoid
function. For suitable parameter regimes, the resulting collective
network dynamics can be chaotic. The presence of inhibitory synapses, however, complicates
the discussion of the learning rule; in particular, there is no good
reason why the learning rules for excitatory and inhibitory synapses
should be the same. As an alternative, one can consider a network with
excitatory synapses only, but with a nonmonotonic dynamical function
for the network elements. Simple such functions are the tent or
logistic maps, which, for a suitable parameter choice, can lead to
chaotic dynamics of the individual elements. Naturally, the collective
network dynamics can then also be chaotic. Here, we choose the second
option, with a logistic map for the individual dynamics. We are thus
in the framework of coupled map lattices as introduced in
\cite{Kaneko84}. In particular, such coupled map lattices can exhibit
synchronized chaotic behavior, see \cite{JJ1} for a systematic
analysis. This is helpful also for understanding some of the weight
dynamics by our STDP rule below.

Our model thus abstracts from many properties of neurobiological
networks, but we hope to thereby identify certain general aspects that
should also be helpful for understanding the structure of 
networks as the result of structural changes from ongoing dynamical
activity. 
We use a coupled map network and a simple learning rule
arising from spike-time-dependent plasticity (STDP). The theoretical
origin of that learning rule is the well known postulate of Hebb
\cite{Hebb} that
 \emph{when an axon of a cell A is near enough to excite cell B or
repeatedly or persistently takes part in firing it, some growth
process or metabolic change takes place in one or both cells such
that A's efficiency, as one of the cells firing B, is increased.}. On the basis
of mathematical considerations\cite{KGvH1,J2} and experimental findings\cite{MLFS,Zh}, a precise
version says that the synaptic strength increases when the presynaptic
neuron fires shortly before the postsynaptic one -- in which case a
causal contribution can be assumed --, but is weakened when that
temporal order is reversed. (This mechanism is called
spike-timing-dependent synaptic plasticity (STDP).) The latter
feature, in an elegant manner, prevents the unbounded
growth in synaptic strengths, a problem with earlier attempts to
implement Hebb's rule. 

Such a mechanism would help to stabilize specific neuronal activity
patterns in the brain. If neuronal activity patterns correspond to
behavior, then the stabilization of specific patterns implies the
learning of specific types of behaviors.

\subsection{Coupled map network}
As already mentioned, our model is not based on the dynamical features
of real neurons that are described by Hodgkin-Huxley type
equations, see e.g. \cite{DA}. We rather consider the
following coupled dynamical system
\begin{eqnarray}\label{eq:cmn}
X_{n+1}=Gf(X_n)
\end{eqnarray}
where $X_n$ is an $N$-dim column vector consisting of the 
state of the nodes of the network, 
$G$ is an $N\times N$ 
coupling matrix that describes how the network is connected, and $f$ is a map from $\Omega=[0,1]^N$ onto
itself modeling the dynamics of the node. We consider chaotic dynamics,  the logistic map defined as:
\begin{eqnarray}\label{eq:map}
f(x) = \mu x (1-x)
\end{eqnarray}
with $\mu=4$. $G_{ij} (i \neq j)$ is the coupling strength of the edge from
$j$ to $i$, and we impose the balancing condition
$G_{ii}=1-\sum_{i\neq j} G_{ij}$, as usual for coupled map lattices
\cite{Kaneko84}.  
$G_{ij}=0$ means that there is no link from $j$ to
$i$. 
We should note that we choose the $G_{ij}$ for $i\ne j$ nonnegative,
that is, we do not consider inhibitory synapses. Dynamically, this is 
compensated by the nonmonotonic behavior of \rf{eq:map}. The key point is that we need to generate some nonmonotonic dynamics. This could be caused either by the interplay of excitation and inhibition in the transmission of activity between elements, as in most neural network models, or by a nonmonotonic dynamics of the elements themselves. We choose the latter option, because then the analysis of synchronization is easier.   
 
The dynamics in equation~(\ref{eq:cmn}) is discrete. The state
vector at time instance $n$ gets updated by the nonlinear map and
the coupling matrix to a new state vector at time instance $n+1$.
Also, the balancing condition mentioned above is imposed at every step.

\subsection{The learning rule}

In our model, the learning rule assigns a dynamics to the entries in the
coupling matrix.
For the above discrete dynamics, the learning rule also has to be
discrete. We choose the following learning rule:
\begin{eqnarray}\label{eq:stdp}
G_{ij}(n+1) = G_{ij}(n) + \epsilon (X_j(n-1)X_i(n) - X_j(n)X_i(n-1))
\mbox{ for }i\ne j,
\end{eqnarray}
where $\epsilon$ is a small parameter deciding the time scale of
the learning dynamics. Thus, the strength of the connection from $j$ to $i$ grows when the state of $j$ at time $n-1$ and the state of $i$ at time $n$ are correlated, and it decreases when the correlation switches the temporal order, that is, when $i$ is active before $j$. 
This rule thus represents a discrete time implementation of STDP. When
two nodes are synchronized the coupling strength does not change. We
recall here that coupled map lattices synchronize under a wide range
of conditions, see \cite{JJ1}, even though synchronization need not
and does not always occur in our simulations.\\
From a different perspective, in connection with information flows in networks, a general class of such learning rules has been considered in \cite{WA}. 

\begin{figure}
\centering
\includegraphics[height=8cm]{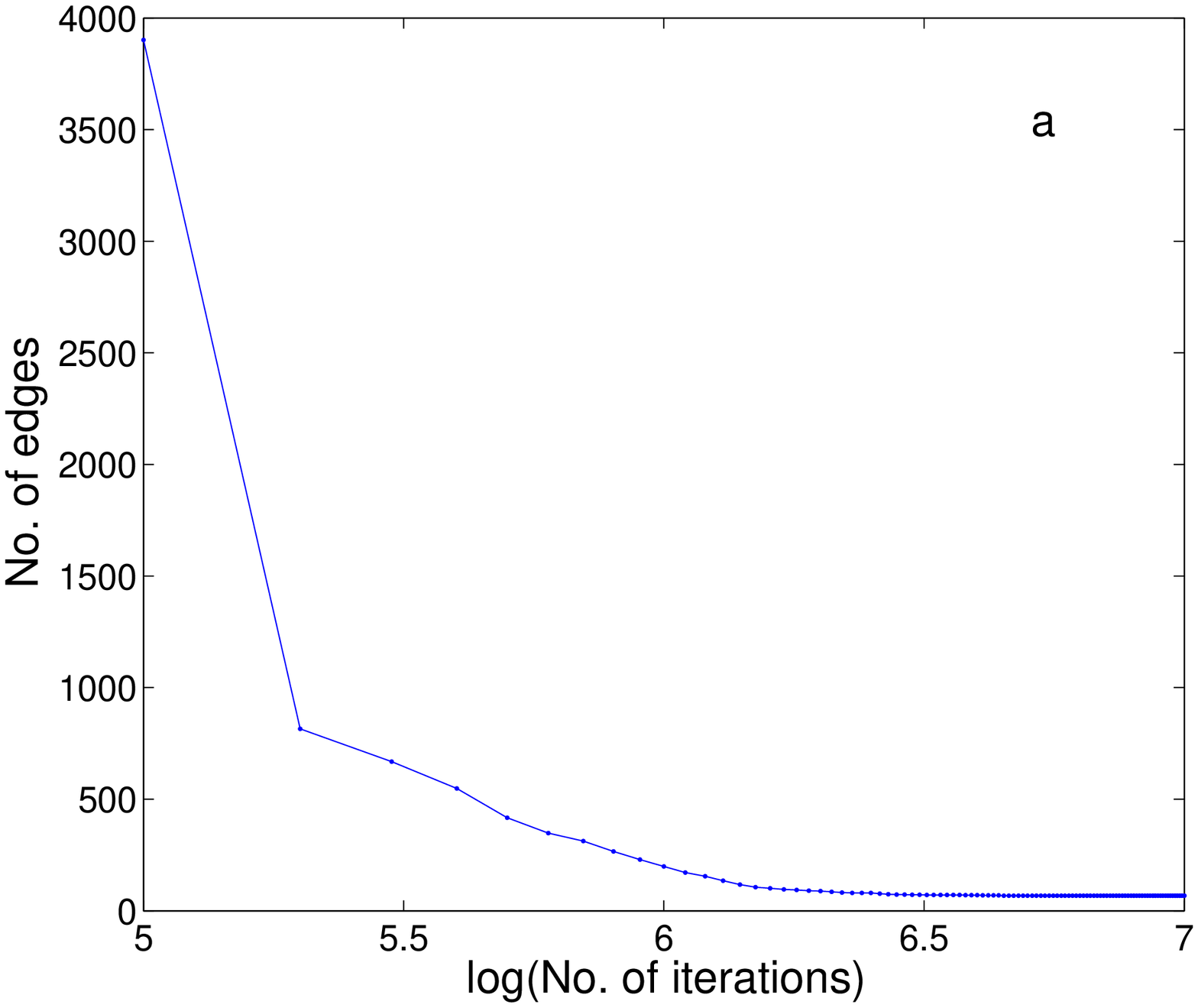}
\includegraphics[height=8cm]{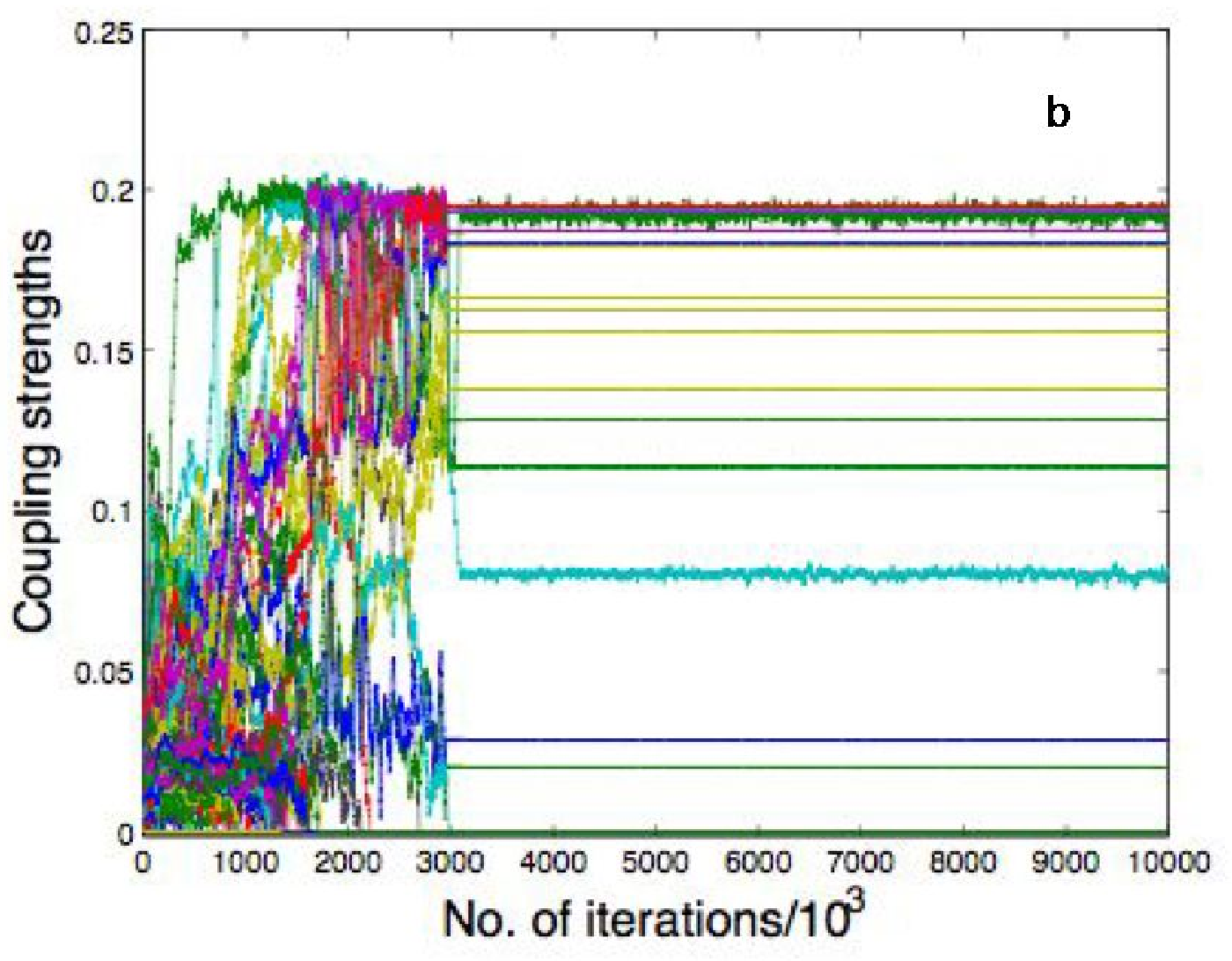}

\caption{ 
(a)The number of edges  plotted against the log of the number of 
iterations. 
(b) Coupling strengths as a function of iterations.
 Number of nodes 
= 64,  $\epsilon = 0.001$ and number of iterations  $=10^7$.}
\label{fig:edges}    
\end{figure}

\subsection{Networks}
Our tools chosen for analyzing our networks are
 motif  and eigenvalue distribution.
Motif distribution refers to the distribution of small subnetworks and
thus yields local information about the network. We consider only
triads and count the number of occurences of different instances
of connected subnetworks containing three sites. There are 13 
different possible connected triads possible. In contrast, 
the eigenvalue distribution of the matrix $G$ yields global information about the network.

\begin{figure}
\centering
\includegraphics[height=8cm]{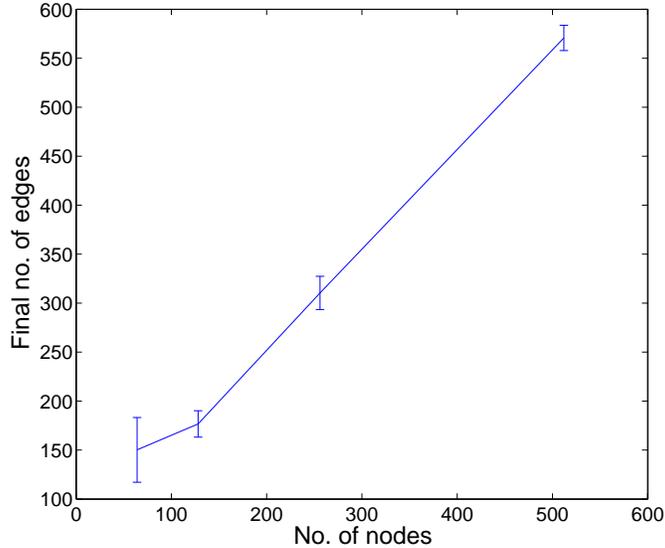}

\caption{Final number of edges as a function of nodes. Error bars give
 standard deviation. Number of realisations = 30, $\epsilon = 0.0001$
and number of iterations = $10^8$.}
\label{fig:fedges}
\end{figure}

\section{Results}
In order to study the effect of learning on the network topology, 
we begin with a globally coupled network of discrete dynamical
systems described in equations~(\ref{eq:cmn}) and (\ref{eq:map}).
The dynamics of the coupling strengths is governed by 
equation~(\ref{eq:stdp}). We assign small nonzero initial
values to the coupling strength and allow the system to evolve for
about $10^7 - 2\times 10^8$ time steps depending on the number of nodes which we have taken 
as powers of 2, from 16 to 2048. Whenever the coupling strength of any edge becomes
negative we clamp it to zero thereafter. This is pruning of the
edge. We find that the evolution leads to a steady state with a robust
heavy tailed network. 

\begin{figure}
\centering
\includegraphics[height=4cm]{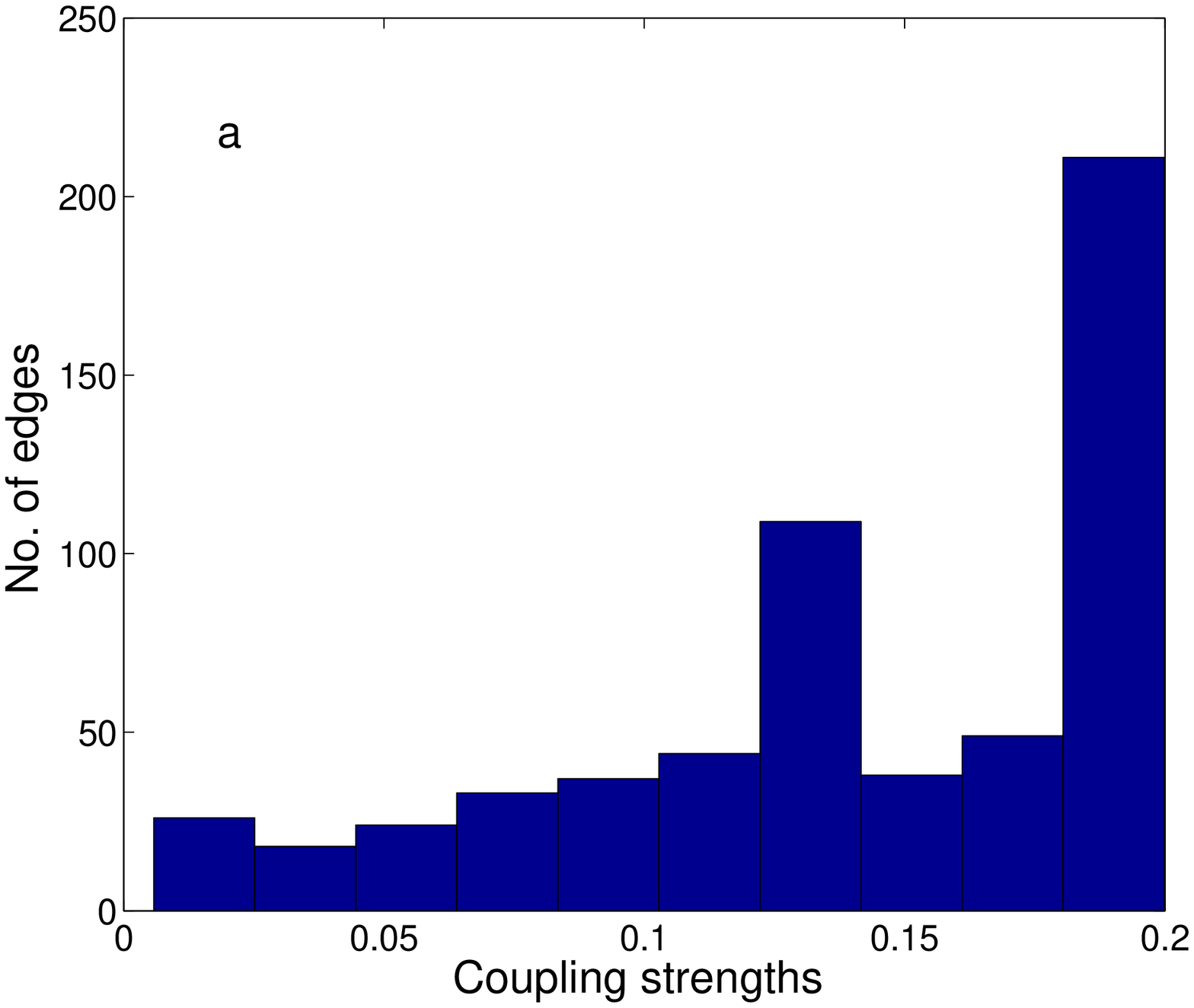}
\includegraphics[height=4cm]{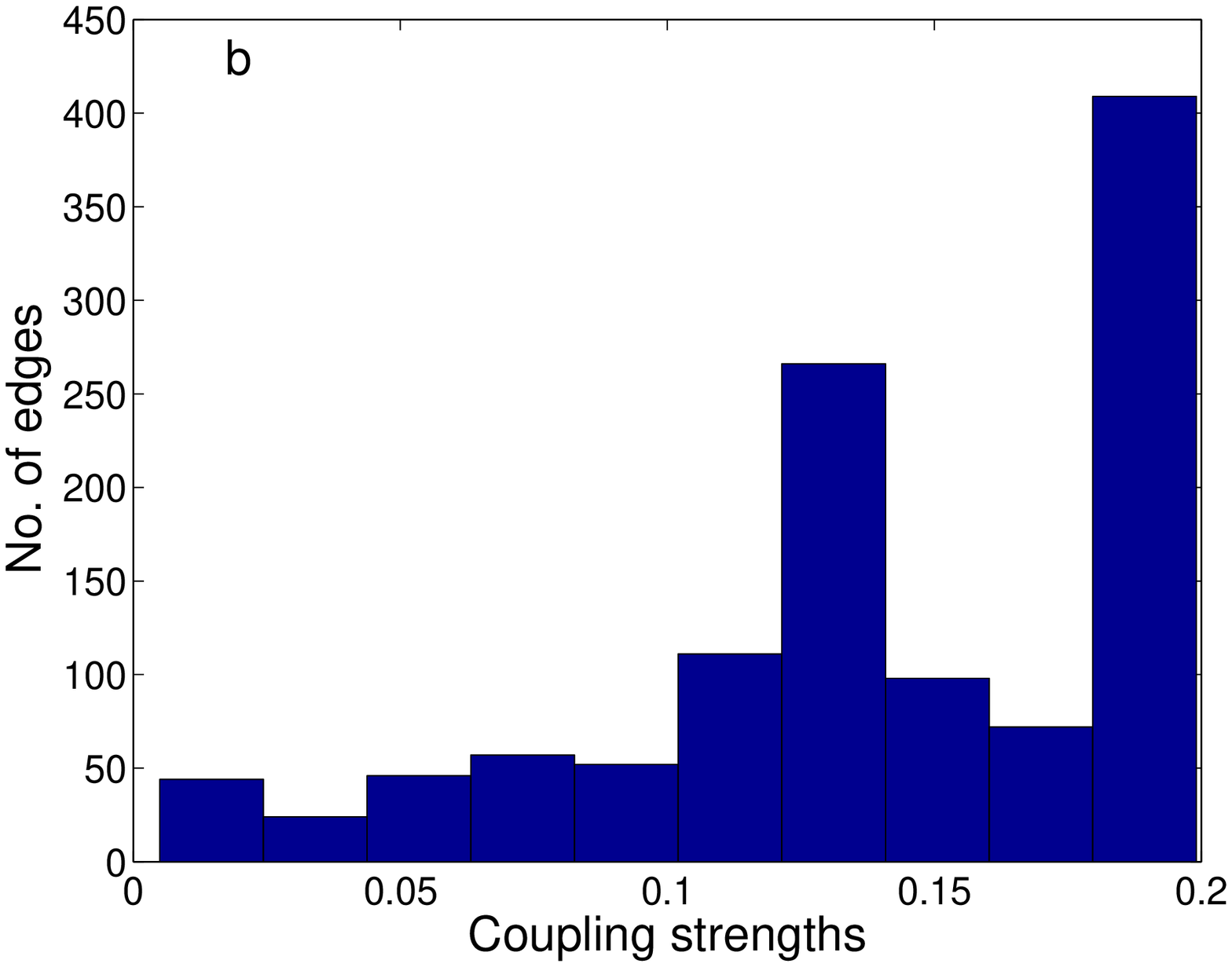}

\caption{Histograms of the coupling strengths of the edges of the residual
network. (a) 512 nodes, $\epsilon = 0.0001$ (b) 1024 nodes, $\epsilon = 0.00005$}
\label{fig:ghist}
\end{figure}

The initial values of strengths of the edges are distributed randomly (and uniformly) in an
interval $[0,g_{in}^{max}]$. The value of $g_{in}^{max}$ is chosen sufficiently small for two
reasons. Firstly, we would like avoid initial near synchronization of any pair of nodes which, we observe,
only slows the approach to the steady state. Secondly, since we start with a globally coupled
network, every node gets an input from many other nodes during the initial phase of the evolution.
If as a result of the addition of these inputs the value of the state variable exceeds unity then
owing to the definition of the logistic map, it will grow without any limit during subsequent
iterations. Therefore we choose $g_{in}^{max} = 0.25 / (N-1)$. Obviously, the value of $\epsilon$
has to be smaller than $g_{in}^{max}$. This severely restricts the choice of the parameter $\epsilon$
in our simulations. This choice of small initial values of the coupling strength is consistent
with the biological example we have in mind. It is natural to expect that, just after  birth,
synapses will be weak and during the course of time some of them will strengthen and others will
be pruned. In any case, only the ratio $g_{in}^{max} /\epsilon$ is  relevant in deciding the
time to reach the steady state.

In Fig.~\ref{fig:edges}a, the number of edges is plotted
against the logarithm of the number of iterations.
We see that the number of edges decreases very fast in the beginning
and then reaches a constant value. 
Fig.~\ref{fig:edges}b shows how individual edges evolve  in a small system. 
The values of coupling strengths
perform a random walk and few of them reach a positive value which is either constant or fluctuates around a
constant mean value.
The strengths of these remaining
edges then are unlikely
to become zero, preserving the structure of the final residual network
intact. We checked this by allowing one realisation to run for $10^8$ iterations. 
We also added a small random noise with uniform distribution to the learning
dynamics. This did not affect the  conclusions. Thus it follows that as the
system evolves the coupling strength of many edges drops to zero but
at the same time some edges become stronger and their strength attains
a steady value. This was found to occur for several values of the system
size as well as different \emph{allowed} values of $\epsilon$.
As depicted in Fig~\ref{fig:fedges}, we found that the number of edges in the
final network is of the order of $N$. The large error bar for a 64 node network could be a result
of the large value of the ratio $g_{in}^{max} /\epsilon$ in this example.

Next we look at the distribution of the coupling strengths in the final
network. The Fig.~\ref{fig:ghist} shows two examples. It is clear that the
final distribution is the same and it is also consistent with the Fig.~\ref{fig:edges}b
which is for a small system.

Having established the existence of a robust network, we can study some of its properties and how robust they are.
First we consider the frequency  of different connected
subnetworks of size three which yields local information
about the network. 
Such a study is thought to be  useful in uncovering the structural design principles
of the network~\cite{Mil}. The network we obtain is rather sparse so we restrict ourselves
only to motifs of size 3.
Out of 13 possible only 4 motifs are present in the final
network irrespective of the value of $\epsilon$ and the network size. If A, B and C are three different nodes then the four subnetworks are:
(1) links going out from (say) B, (2) 
opposite of (1), i.e., links coming into B, (3) there is a link from 
A to B and a link from
B to C and (4) a cyclic triangle. The other
configurations are absent. In particular,  triads with  double links, i.e., a link from A to B and also
from B to A, are absent. This is expected from the learning rule 
since $G_{ij}+G_{ji}$ is a constant and so,  one of the link grows at the
expense of the other. Noncyclic triangles
are also absent.

\begin{figure}
\centering
\includegraphics[height=4cm]{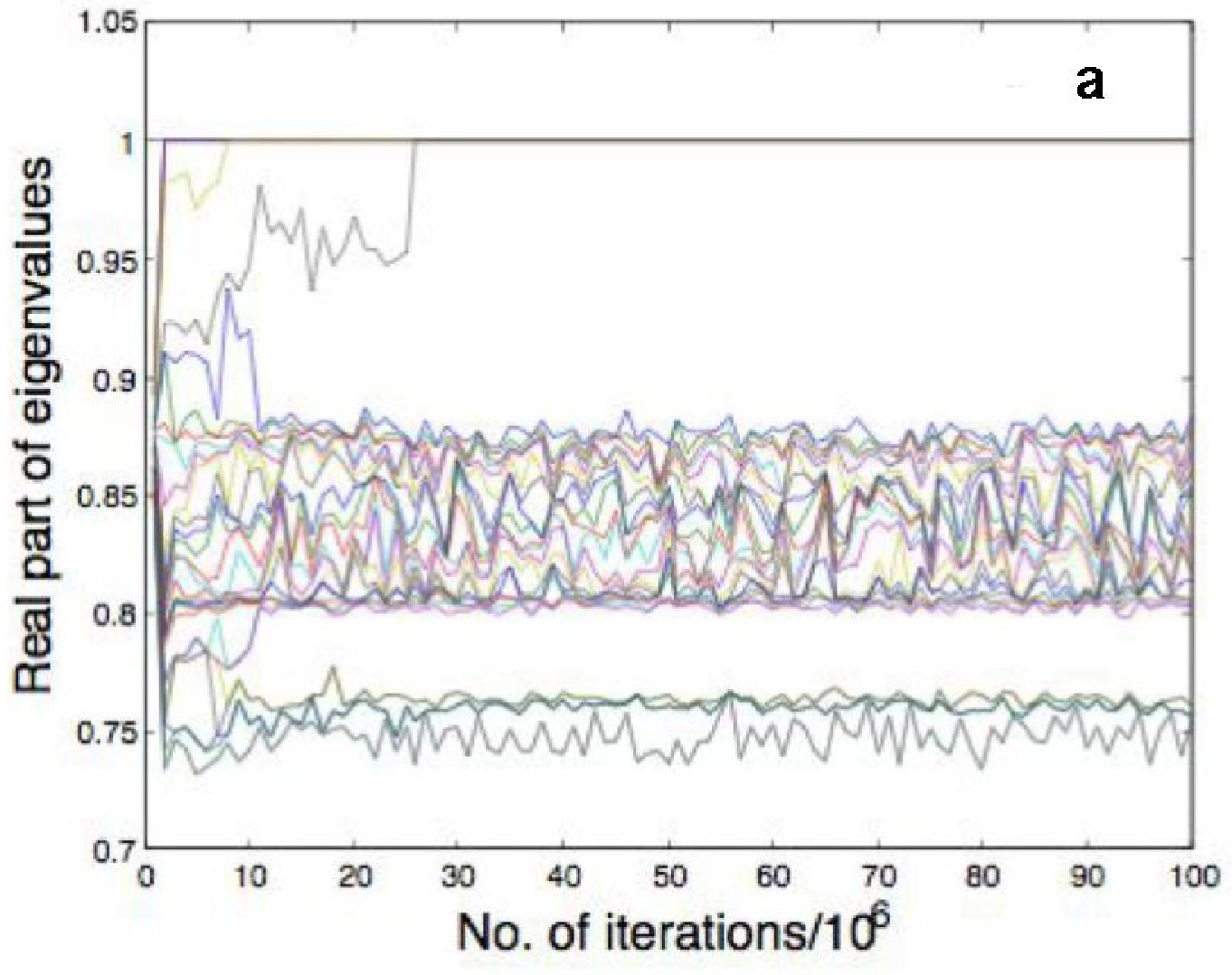}
\includegraphics[height=4cm]{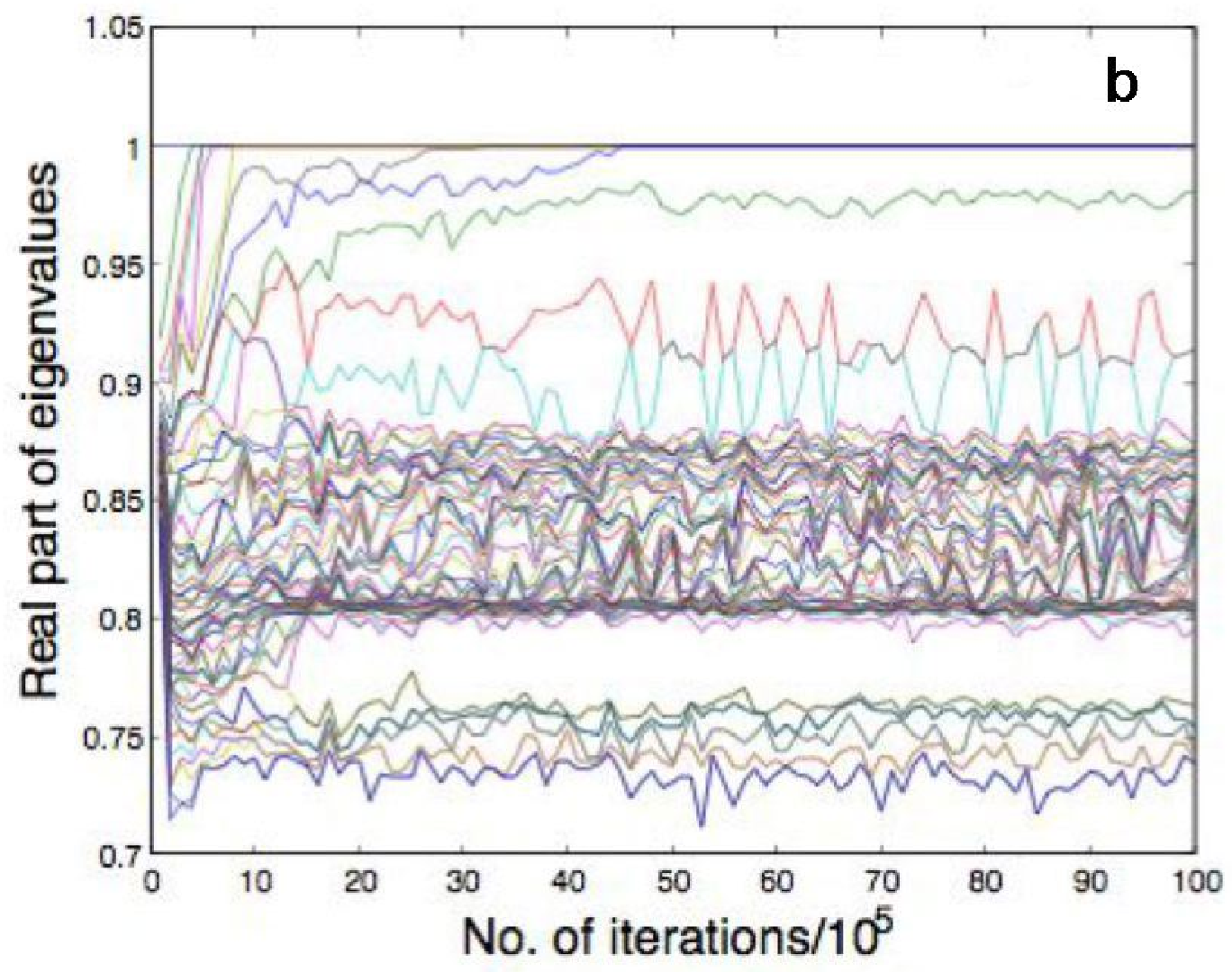}

\caption{This figure shows how the real parts of the eigenvalues
evolve. Number of nodes (a) 32 (b) 64.}  
\label{fig:eig32}

\end{figure}

\begin{figure}
\centering
\includegraphics[height=8cm]{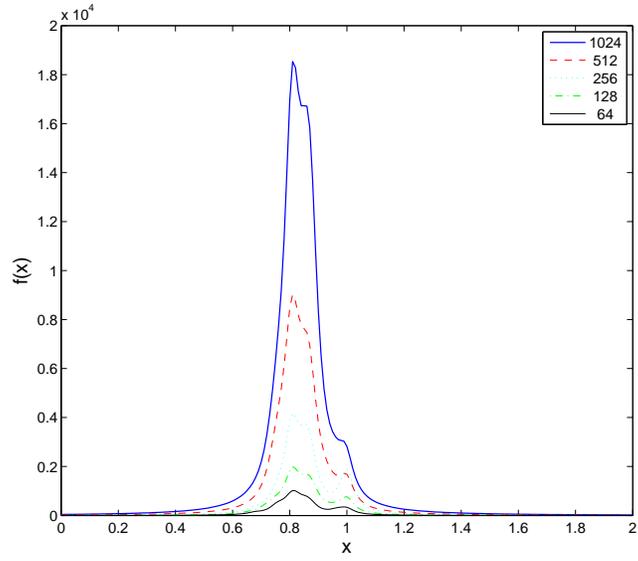}
\caption{Convolution of the real part of the eigenvalues for network size from 64 to 1024.}  
\label{fig:spplot_all}
\end{figure}

Now we study the eigenvalue distribution of the  matrix $G$ which
yields global information about the network. In Figs.~\ref{fig:eig32}a
and~b we plot the real part of the eigenvalues for a 32-node
network and a 64-node network respectively. We see that the real part
of a 
few eigenvalues becomes 1 corresponding to few nodes getting isolated.
There is a lower bound which the real part of eigenvalues do not cross.
This is a consequence of the existence of triangles.
All other eigenvalues reach a steady value which is intermediate.
In fact, they separate
into bands which are robust. In order to succinctly show this for
several system sizes we plot,
in Fig.~\ref{fig:spplot_all},
the convolution of the real parts of the eigenvalues
with Lorentz kernel as given by the function
\begin{eqnarray}\label{eq:spplot}
f(x) = \sum_{j}{\gamma\over{(\lambda_j - x)^2+\gamma^2}}.
\end{eqnarray}  
where $\lambda_j$s are the real parts of the eigenvalues of the matrix
$G$ and $\gamma = 0.03$.
The similar structure of the graphs over several system sizes is evident.

\begin{figure}
\centering
\includegraphics[height=8cm]{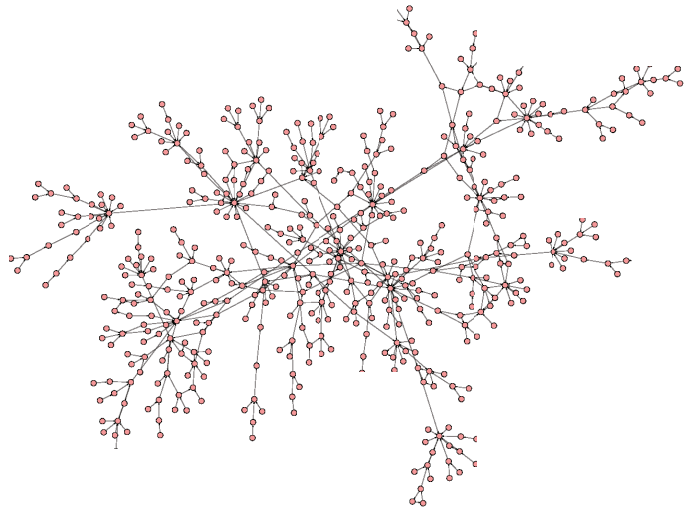} 
\caption{The largest component of the final network of 1024 nodes after $2 \times 10^8$ iterations.
$\epsilon=0.00005$ }   
\label{fig:graph}
\end{figure}

\begin{figure}
\centering
\includegraphics[height=6cm]{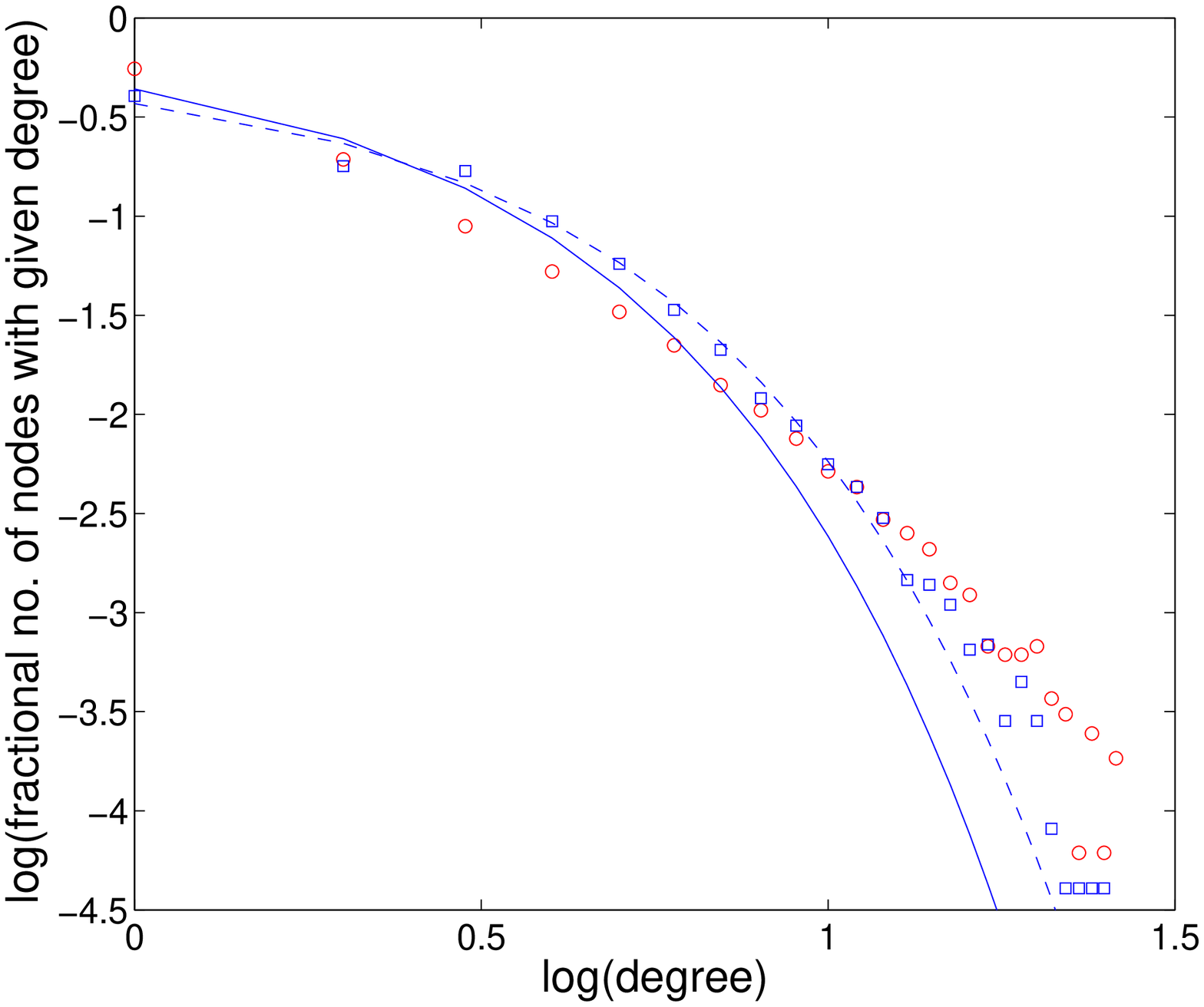} 
\caption{The degree distribution for networks of 512 (blue squares) 
averaged over 50 realizations and 1024 (red circles) nodes averaged 
over 16 realizations. The geometric distributions with same mean are
also plotted for comparison. The continuous line is for 1024 nodes
and the dashed line is for 512 nodes. }   
\label{fig:power}
\end{figure}

\begin{figure}
\centering
\includegraphics[height=6cm]{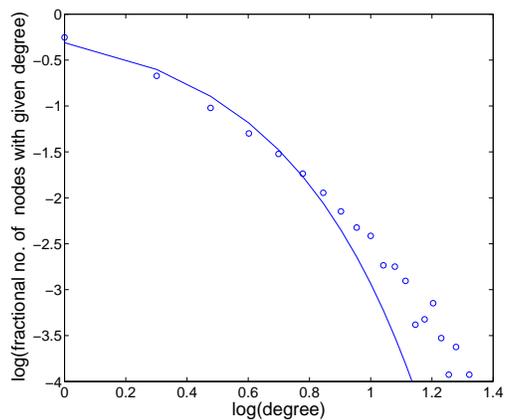}
\caption{The degree distribution for networks of 2048 nodes after $2 \times 10^7$ iterations
averaged over 9 realizations. 
The geometric distribution with same mean is
also plotted for comparison. 
}
\label{fig:powerl}
\end{figure}

\begin{figure}
\centering
\includegraphics[height=6cm]{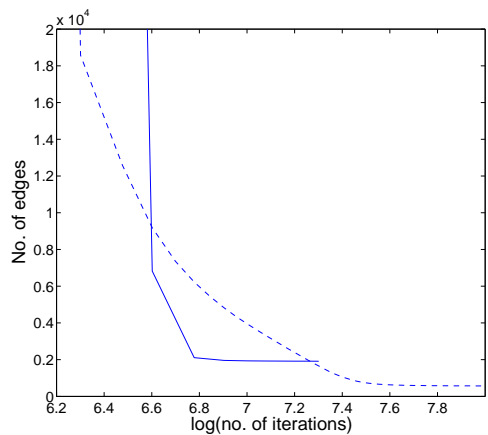}
\caption{ The average number of edges  plotted as a function of the logarithm of
the number of iterations. The continuous line corresponds to a 2048 node network of
Fig.~\ref{fig:powerl} and the dashed line is for a 512 node networks in Fig.~\ref{fig:power} 
}
\label{fig:edgesl}
\end{figure}

The most interesting outcome of this study is the final structure
of the graph. We find that though some nodes and small clusters get
separated there is still a single large connected component. In
Fig.~\ref{fig:graph} we show the residual graph of 1024 nodes.
The network has broad degree distribution. 
In Fig.~\ref{fig:power}, we show the degree distribution for 512 and 1024
node networks. We compare it with a geometric distribution with the same
mean. The deviation from the exponential decay is clear. 
We have also started from a initial random network with degree 100
instead of the globally coupled network. This system reaches a steady
state in less number of iterations and allows us to increase the size
of the system. The degree distribution arising
out of this simulations for 2048 node network is shown in Fig.~\ref{fig:powerl} along with
corresponding geometric distribution. This shows that the choice of
initial network is not crucial for the final conclusion.
In all these simulations we have made sure that the system has
reached a steady state by keeping track of the number of edges. Fig.~\ref{fig:edgesl} 
displays this fact.

\begin{figure}
\centering
\includegraphics[height=6cm]{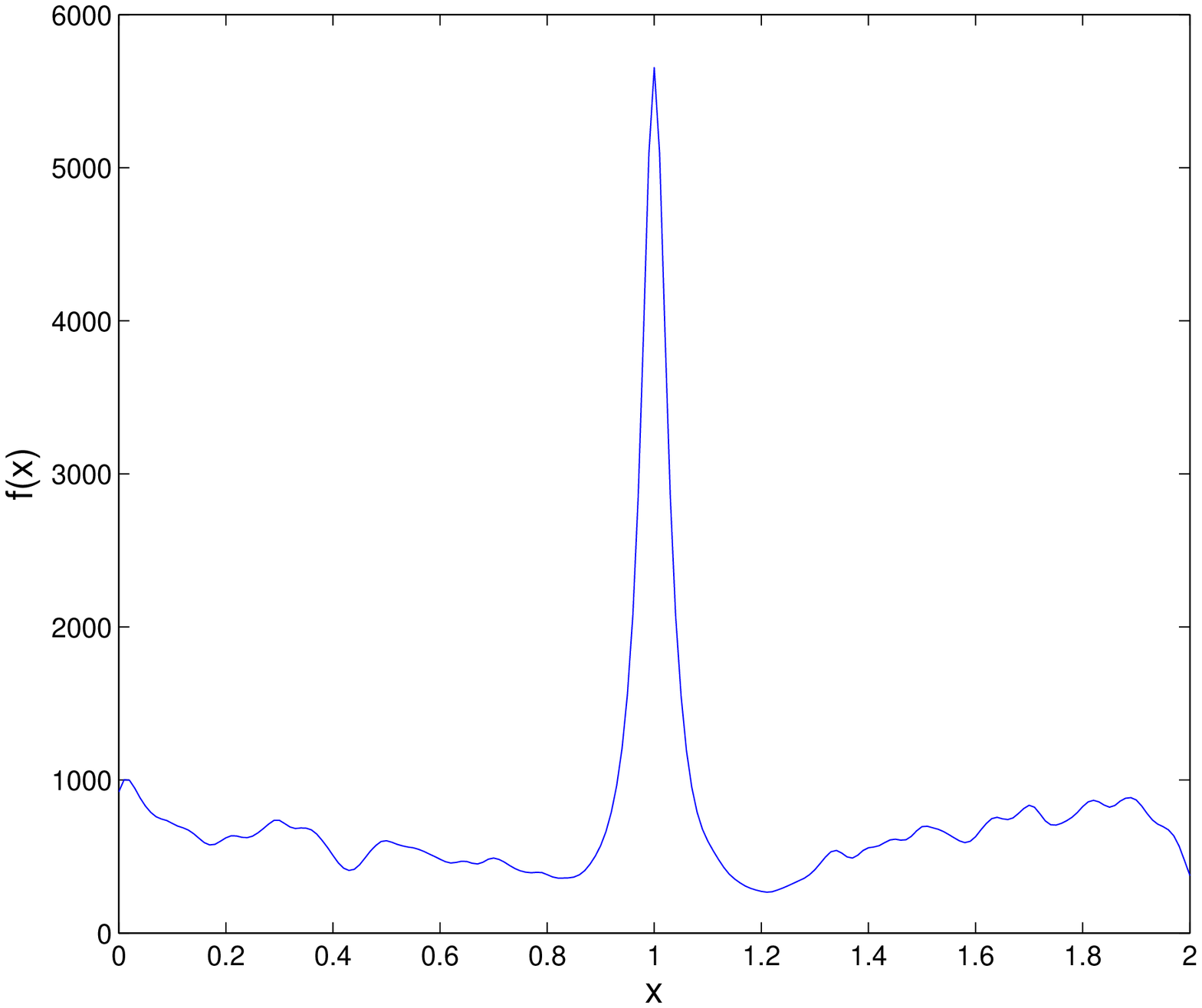} 
\caption{The spectral plot of the Laplacian of the underlying undirected
graph for 1024 node network}   
\label{fig:spl}
\end{figure}

Now we study the structure of the underlying network. That is, we
consider the network disregarding the directions of the edges and 
study the eigenvalues of the Laplacian defined by
\begin{eqnarray}
\Delta v(i) := v(i) - {1\over n_i}\sum_{j,j\sim i} v(j).
\end{eqnarray} 
where $v(i)$ is the state of site $i$ and $n_i$ is the number of neighbors of site $i$.
$j\sim i$ implies that $i$ and $j$ are neighbors.
The spectrum of this operator yields important invariants of the underlying
graph, see \cite{BJ1}.
We again plot the spectral plot as in equation~(\ref{eq:spplot}) of
the eigenvalues of this operator in Fig~\ref{fig:spl}. The structure
of this plot is somewhat different from the ones obtained for other
networks, see \cite{BJ2} for a systematic comparison with theoretical
paradigms and empirical networks,  even though the
prominent peak near 1 is a feature exhibited by many empirical
networks, see the diverse examples in \cite{BJ2}. 

\section{Discussion}
We have studied the effect of a STDP type learning rule on the network
dynamics. This learning rule incorporated  
the necessary competition between different edges.
As the network evolves, some edges grow in strength and some
edges become weak and are then eliminated.  
This is much like the biological case wherein the brain is 
densely wired at the time of  birth and then most synapses are pruned
in the course of the development. 
In our setting, the resulting network is sparse and  has broad degree distribution. 
The properties of the final network are universal in the sense
that they do not depend on the system size and the value of  $\epsilon$
within the allowed range. This suggests that there
exists an attractor of the overall dynamics (nodes, synapses and network)
to which the system evolves.
This constitutes a mechanism to obtain heavy-tailed networks where no 
preferential attachment has been explicitly introduced. 
There exist some other works that obtain scale-free networks without
overt preferential attachment, for example,~\cite{BM,EB,Rol}.
The network we obtain is rather
too sparse as compared to the real biological examples. The reason for this seems 
to be that our simple model only incorporates pruning due to learning dynamics. There are
many processes, like exuberance, that can introduce new links. It
remains to study the effects of such or other, biologically more realistic, 
dynamics and learning rules.

\vspace{0.1in}
\noindent
\emph{Note:} After we had communicated this manuscript we became aware of the work
by Shin and Kim~\cite{SK} in which they have obtained similar results using 
the FitzHugh-Nagumo model.

\section*{Acknowledgement}
KMK would like to acknowledge the financial assistance from DST
(India). We should also thank the referees for their useful comments.

\end{document}